\documentclass[floatfix,showpacs,showkeys,twocolumn,prl]{revtex4}
\usepackage{amssymb}
\usepackage{amsmath}
\usepackage{bm}
\usepackage{epsfig}
\usepackage{comment}
\usepackage{graphicx}
\newcommand{\be}{\begin{eqnarray}} 
\newcommand{\ee}{\end{eqnarray}} 
\newcommand{\bml}{\begin{multline}}
\newcommand{\eml}{\end{multline}}

\begin{document}
\title{Physical condition and spin-resolved exchange correlation kernels\\
 in an inhomogeneous many electron system}
\author{Ina Yeo}
\author{Kyung-Soo Yi}
\affiliation{Department of Physics\\
and\\
Research Center for Dielectric and Advanced Matter Physics\\
Pusan National University, Busan 609-735, Korea}
\date{today}

\begin{abstract}
We first exploit the spin symmetry relation 
$f^{\rm xc}_{s\bar s}(\zeta)=f^{\rm xc}_{\bar s s}(-\zeta)$  
for the exact exchange correlation kernel $f^{\rm xc}_{s\bar s}(\zeta)$
in an inhomogeneous many electron system with arbitrary spin polarization $\zeta$.
The physical condition required to satisfy the specific symmetry relation 
$f^{\rm xc}_{s\bar s}(\zeta)=f^{\rm xc}_{\bar s s}(\zeta)$ is derived
and examined for simple ferromagnetic-nonmagnetic structure by taking the electrochemical potential into account. 
The condition is then applied to several composite systems useful in spintronics applications such as the magnetic 
system with net spin polarization. 
\end{abstract}

\pacs{72.15.Gd, 71.45.Gm, 73.21.-b, 72.10.Fk}

\keywords{Exchange correlation kernel, 
Spin symmetry relation, Electrochemical potential, Spin injection}

\maketitle

Recently the spin transport properties have been attracting great interest 
for their potential applications to spintronics and quantum computation
\cite{wolf,Yu,Kikkawa,kimura1}.
One of the central problems in these fields is controlling the distribution of spin-polarized carriers 
in multicomponent structure alternating ferromagnetic and nonmagetic materials.
In this system, the exchange correlation (XC) kernel can be useful for giving direct insight 
into the carrier population in spin-polarized channels.  
Exchange correlation kernel (XCK) $f^{\rm xc}_{ss'}(\zeta)$ is defined by 
\begin{eqnarray}
f^{\rm xc}_{ss'}(\zeta)\equiv
\frac{\partial^2 E^{\rm xc}(\boldsymbol r,\boldsymbol r';\zeta)}
{\partial n_{s'}(\boldsymbol r)\partial n_{s}(\boldsymbol r')}
\end{eqnarray}
where $E^{\rm xc}$ is the total XC energy functional in many electron system with spin polarization $\zeta$.
$f^{\rm xc}_{ss'}(\zeta)$ is a basic
concept in describing many body correlation effects in an inhomogeneous electron liquid
and satisfies the symmetry relation $f^{\rm xc}_{s\bar s}(\zeta)=f^{\rm xc}_{\bar s s}(-\zeta)$
where $s(\bar s)$ refers to the majority (minority) spin.
The spin symmetry relation of XCK 
plays a significant role to understand the nature of the spin-spin response function 
(which contains spin symmetric and anti-symmetric parts of XCK) in artificial composite structure. 
Unfortunately, $E^{\rm xc}$, a key ingradient of XCK,
is not capable of providing accurate
$f^{\rm xc}_{ss'}$ in spite of numerous studies including density gradient corrections \cite{Giuliani,perdew,phil}.
They fail to reduce the mean absolute error to the desired level in the chemical bonding energies. 
That means that the desired chemical accuracy has not been reached yet.
While the ``mixed scheme'' \cite{roi,julien,pazini,fran} combining the density functional theory (DFT)
with other methods such as quantum Monte Carlo simulations and coupled cluster
calculations applied separately to the short and long range parts of the electron-electron interaction 
has been proposed as an alternative, 
the short range part of the Coulomb interaction between electrons is still well described 
by semilocal functionals \cite{pazini,fran}.
On the other hand, gradient corrected density functionals (GCDF) \cite{Giuliani,perdew,phil} have been
used for studies of electronic structures but mainly been restricted to unpolarized systems. Calculation 
of density functionals in the spin-polarized system has been extended to
the local spin density functional scheme combining with other approximation methods \cite{pazini}.
Extension of GCDF to the spin-polarized system is not available yet.
Hence the symmetry relation of the ``exact'' spin-resolved XCK has not been well established
in an inhomogeneous spin polarized system.
In this paper, we first exploit the symmetry relation of the ``exact'' spin-resolved XCK
and the condition required to satisfy the specific symmetry relation (SSR)
$f^{\rm xc}_{s\bar s}(\zeta)=f^{\rm xc}_{\bar s s}(\zeta)$ of the exact XCK.\

Although considerable researches have been devoted to investigate spin current
$I_{s(\bar s)}$, there has been no attempt to interpret XCK  by directly measurable quantities
such as 
$I_{s(\bar s)}\propto-\frac{\partial\mu_{s(\bar s)}}{\partial
x}$ where $\mu_{s(\bar s)}$ is the electrochemical potential (ECP) in spintronics.
In view of the fact that the required condition to satisfy SSR could be 
related to the spin density variation $\nabla n_{s(\bar s)}$
and then $\nabla n_{s(\bar s)}$ to the ECP variation $\nabla \mu_{s(\bar s)}$, 
information on a system can be obtained straight by experimental observations of spin related phenomena. 
Also, theoretical ECP can exactly be checked through the symmetry relation of XCK.
Hence we first propose the proper situations satisfying SSR in alternating multilayer system.
We also give the accurate relation of corresponding spin-resolved pair
correlation functions $g_{ss'}(\boldsymbol r)$ and $g_{ss'}(\boldsymbol r')$.
 
GCDF is given by the sum
of the kinetic energy $T$ of a noninteracting particle system, potential energy $U$, 
and unknown functional $E^{\rm xc}$
\begin{eqnarray}
E[n(\boldsymbol r)]=T[n(\boldsymbol r)]+U[n(\boldsymbol r)]+E^{\rm xc},
\end{eqnarray}
where one particle density $n(\boldsymbol r)$ is written by
\begin{eqnarray}
n(\boldsymbol r)=N\sum_{s_2,\cdots,s_N}\int\lvert\Psi(\boldsymbol r_1 s_1,\cdots,\boldsymbol r_N s_N)\lvert^2
d\boldsymbol r_2,\cdots,d\boldsymbol r_N
\end{eqnarray}
with spins of $N$ electrons. 

In pair density theory giving more accurate value of the ground state energy 
than one particle density \cite{Gonis}, 
the spin-summed pair density is given by
$n(\boldsymbol r_1,\boldsymbol r_2)=\frac{N(N-1)}{2}\sum_{s_1 s_2}\gamma_{s_1 s_2}(\boldsymbol r_1,\boldsymbol r_2)$.
Here $\gamma_{s_1 s_2}(\boldsymbol r_1,\boldsymbol r_2)$ is the spin-resolved diagonal 
of the two-body reduced density matrix \cite{Gori}
\begin{align}
\gamma_{s_1 s_2}(\boldsymbol r_1,\boldsymbol r_2)&  \nonumber \\ 
=\sum_{s_3,\cdots,s_N}
&\int\lvert\Psi(\boldsymbol r_1 s_1,\cdots,\boldsymbol r_N s_N)\lvert^2d\boldsymbol r_3,\cdots,d\boldsymbol r_N.
\label{PLR}
\end{align}
The exact energy density functional is given by $E[n(\vec x)]=T[n(\vec x)]+U$ with
$\vec x=(\boldsymbol r_{1},\boldsymbol r_{2})$.
Here, $U$ consists of the external potential of a given pair and 
the interaction potential between particles forming two pairs at 
$\vec x_i=(\boldsymbol r_{i1}, \boldsymbol r_{i2})$ 
and $\vec x_j=(\boldsymbol r_{j1},\boldsymbol r_{j2})$ \cite{Gonis}.
Hence, the total XC energy functional $E^{\rm xc}$ in interacting system
can be described in terms of pair density $n(\vec x)$
\begin{eqnarray}
E^{\rm xc}[n(\boldsymbol r)]=T[n(\vec x)]-T[n(\boldsymbol r)]+U[n(\vec x)]-U[n(\boldsymbol r)].
\end{eqnarray} 
 $E^{\rm xc}$ contains the correlated kinetic term 
$T^{\rm xc}\equiv T[n(\vec x)]-T[n(\boldsymbol r)]$ as well as the interparticle interaction potential.
Here, $T^{\rm xc}$ denotes the difference between the exact kinetic term $T[n(\vec x)]$ on interacting scheme
and noninteracting counter part $T[n(\boldsymbol r)]$.  
$f^{\rm xc}_{s\bar s}$,  
the second derivatives of $E^{\rm xc}$ with respect to spin densities $n_{s}(\boldsymbol r)$ 
and $n_{\bar s}(\boldsymbol r')$
at a given pair position $\vec x_i=(\boldsymbol r,\boldsymbol r')$,
is now written by 
\begin{eqnarray}
f^{\rm xc}_{s\bar s}(\zeta)
=\frac{\nabla_{\boldsymbol r}\nabla_{\boldsymbol r'}[T^{\rm xc}[n(\vec x)]+U[n(\vec x)]-U[n(\boldsymbol r)]]}
{\nabla n_{\bar s}(\boldsymbol r) \nabla n_{s}(\boldsymbol r')}.
\end{eqnarray}
For the exactly defined XCK, the symmetry relation  
$f^{\rm xc}_{s\bar s}(\zeta)=f^{\rm xc}_{\bar s s}(-\zeta)$ 
is trivial in an inhomogeneous broken spin symmetry system
since
majority and minority spins interchange their orientations and positions 
with the reversed polarization $-\zeta$.
That is, based on the assumption that the physical condition
\begin{eqnarray}
\frac{\nabla n_{\bar s}(\boldsymbol r')}{\nabla n_{s}(\boldsymbol r')}=
\frac{\nabla n_{\bar s}(\boldsymbol r)}{\nabla n_{s}(\boldsymbol r)}\label{key}
\end{eqnarray}
is fulfilled (i.e., the ratios of spin density gradients
are the same at two different positions $\boldsymbol r$ and $\boldsymbol r'$),
SSR 
\begin{eqnarray}
f^{\rm xc}_{s\bar s}(\zeta)=\frac{\partial^2 E^{\rm xc}(\boldsymbol r,\boldsymbol r';\zeta)}
{\partial n_{\bar s}(\boldsymbol r)\partial n_{s}(\boldsymbol r')}
=\frac{\partial^2 E^{\rm xc}(\boldsymbol r,\boldsymbol r';\zeta)}
{\partial n_{s}(\boldsymbol r)\partial n_{\bar s}(\boldsymbol r')}=f^{\rm xc}_{\bar s s}(\zeta)\label{syr}
\end{eqnarray}
is obtained trivially.
In other words, only ratios of spin density variations at different sites 
are required to investigate SSR.
This condition is valid in various density varying systems.
 By examining the condition given by Eq.(\ref{key}) in various spin valve systems,
we can investigate the validity of SSR. 
Given the relation between ECP $\mu_{s(\bar s)}$ 
and the nonequilibrium spin carrier density $n_{s(\bar s)}$ 
in metallic and nonmetallic system, 
spin density variation $\nabla n_{s(\bar s)}$ can be obtained. 

 In a highly degenerate system, the density variations for spin-up and spin-down carriers are given,
 in the presence of an electric field $\boldsymbol E=-\nabla{\Phi}$, by \cite{Yu}
\begin{eqnarray}
\nabla n_{s(\bar s)}=e\nabla D_{s(\bar s)}(\epsilon_{\rm F})[\mu_{s(\bar s)}+e\Phi]+
eD_{s(\bar s)}(\epsilon_{\rm F})[\nabla\mu_{s(\bar s)}-e\boldsymbol E],\nonumber
\end{eqnarray}
where $D_{s(\bar s)}(\epsilon_{\rm F})$ is the spin-up (spin-down) density of states 
at the Fermi level.
Depending on the dimension of multilayer structure, $D_{s(\bar s)}$ is varied
but the gradient of $D_{s(\bar s)}(\epsilon_{\rm F})$ at the fixed Fermi level vanishes always.
Hence the condition (\ref{key}) can be written by 
\begin{eqnarray}
\frac{D_{s}(\epsilon_{\rm F})}{D_{\bar s}(\epsilon_{\rm F})}\cdot
\frac{\nabla\mu_{s}-e\boldsymbol E}{\nabla\mu_{\bar s}-e\boldsymbol E}\Bigg\lvert_{\boldsymbol r}
=\frac{D_{s}(\epsilon_{\rm F})}{D_{\bar s}(\epsilon_{\rm F})}\cdot
\frac{\nabla\mu_{s}-e\boldsymbol E}{\nabla\mu_{\bar s}-e\boldsymbol E}\Bigg\lvert_{\boldsymbol r'}
.\label{condi}
\end{eqnarray}
In nonmetallic region of a homogemeous system with no space charge,
the ratios of local variances $\nabla n_s /\nabla n_{\bar s}$ are the same.
In doped systems, spin polarization can be created 
keeping the total number of electrons and holes constant, 
$\delta n_s+\delta n_{\bar s}=0$ \cite{Yu}.
For a constant equilibrium spin density $n^0$, it is a trivial situation satisfying SSR since
$\frac{\nabla n^0_s+\nabla\delta n_s}{\nabla n^0_{\bar s}+\nabla\delta n_{\bar s}}\Big\lvert_{\boldsymbol r}
=-1$.

We first consider the general ECP $\mu_{s(\bar s)}$ \cite{valet} in alternating ferromagnets (F) and nonmagnets (N) 
to see in what region SSR is satisfied.
In a homogeneous F with ``up'' magnetization, 
the local variation of $\mu_{s(\bar s)}$ with respect
to $x$ perpendicular to the layer is derived, at zero temperature, by
\begin{eqnarray}
\frac{\partial \mu_{s(\bar s)}}{\partial
  x}=e E+\frac{(\beta\pm1)}{l_{sf}^{\rm F}}
  [K^{(n)}_2e^{x/l_{sf}^{\rm F}}
  -K^{(n)}_3e^{-x/l_{sf}^{\rm F}}]\label{gef}
  \end{eqnarray}
where $\beta$ and $l_{sf}^{\rm F}$ are the bulk spin asymmetry coefficient
$(-1<\beta<1)$ and spin diffusion length, respectively, in F.
The constants $K^{(n)}_i$ are determined from the proper
boundary conditions in the $n$th layer.
For a given bulk spin resistivity $\rho_{s(\bar s)}=1/\sigma_{s(\bar s)}=2\rho^*_{\rm F}[1\mp\beta]$
in F, $\beta$ can be determined from the spin-scattering measurements \cite{valet,vouille}.
Here, $\rho^*_{\rm F}$ is the fixed resistivity in F obtained by measurement
and $-(+)$ sign in front of $\beta$ corresponds to the up (down) spin. 
Near the interface, scattering is localized significantly in an interfacial region.
Since the spin flip scattering rate is not easily manipulated and identified near the interface,
obtaining accurate $\beta$ is not easy.
As the stabilization of $\beta$ is determined by the impurity concentration,
$\beta$ should be checked with care at positions $\boldsymbol r$ and $\boldsymbol r'$ in a region especially near the interface.
From Eqs.(\ref{key}) and (\ref{condi}), the condition required to satisfy SSR is 
\begin{eqnarray}  
\frac{\beta+1}{\beta-1}\bigg\lvert_{\boldsymbol r}=\frac{\beta+1}{\beta-1}\bigg\lvert_{\boldsymbol r'}
  &&   (i.e., \frac{l_{sf}}{l_{{\bar s}f}}\bigg\lvert_{\boldsymbol r}
=\frac{l_{sf}}{l_{{\bar s}f}}\bigg\lvert_{\boldsymbol r'}).
\end{eqnarray}

For a homogeneous N with ``up'' magnetization, the local variation of the 
general ECP in the $n$th layer is derived, similarly, by 
\begin{equation}
\frac{\partial \mu_{s(\bar s)}}{\partial
  x}=eE\pm\frac{1}{l_{sf}^{\rm N}}[K^{(n)}_2e^{x/l_{sf}^{\rm N}}
  -K^{(n)}_3e^{-x/l_{sf}^{\rm N}}].\label{gen}
\end{equation}
The ratio of spin density variations is simply $-\frac{D_{s}(\epsilon_{\rm F})}{D_{\bar s}(\epsilon_{\rm F})}$
and SSR is trivially satisfied.

Based on the results of the general cases given by Eqs.(\ref{gef}) and (\ref{gen}), we consider a simple F/N structure
as schematically shown in Fig.1(a).
A spin-polarized current with density
$j_{s(\bar s)}$ flows from F ($x<0$) into N ($x>0$) along the
direction perpendicular to the interface.
$\mu_{s(\bar s)}$ 
can be expressed in terms of the spin-resolved conductivity $\sigma_{s(\bar s)}$ and
current density $j_{s(\bar s)}$ 
with the spin accumulation balanced by the spin flip scattering.
 The local variation of ECP is written by
$\frac{\partial\mu_{s(\bar s)} }{\partial
  x}=-2e\rho^*_{\rm F}[1\mp\beta]j_{s(\bar s)}$ in F 
  and 
  $\frac{\partial\mu_{s(\bar s)}}{\partial
  x}=-2e\rho^*_{\rm N}j_{s(\bar s)}$ in N
where $\rho^*_{\rm N}$ is the resistivity of N \cite{son}.
In this case, the physical condition, Eq.(\ref{key})
becomes
\begin{equation}
\frac{(1-\beta)j_{s}+\Upsilon E}{(1+\beta)j_{\bar s}+\Upsilon E}\bigg\lvert_{\boldsymbol r}
=\frac{(1-\beta)j_{s}+\Upsilon E}{(1+\beta)j_{\bar s}+\Upsilon E}\bigg\lvert_{\boldsymbol r'}\label{simple}
\end{equation} 
in F 
and 
\begin{equation}
\frac{j_{s}+\Upsilon E}{j_{\bar s}+\Upsilon E}\bigg\lvert_{\boldsymbol r}
=\frac{j_{s}+\Upsilon E}{j_{\bar s}+\Upsilon E}\bigg\lvert_{\boldsymbol r'}\label{simple1}
\end{equation} 
in N with $\Upsilon=1/(2\rho^*_i)$, ($i$=F, N). 
 Far from the interface ($x\gg 0$) in N with equilibrium spin
state, $\sigma_s=\sigma_{\bar s}$ and $j_s=j_{\bar s}$ \cite{son}.   
This is an obvious situation satisfying SSR 
since $\nabla n_{s}/\nabla n_{\bar s}$ is constant
in the region $x\gg0$.
Near the interface with $\mu_{s}-\mu_{\bar s}\neq0$ (due to the spin scattering),  
the conditions (\ref{simple}) and (\ref{simple1}) should be checked with care in order to satisfy SSR.
\begin{figure}
\begin{center}
\includegraphics{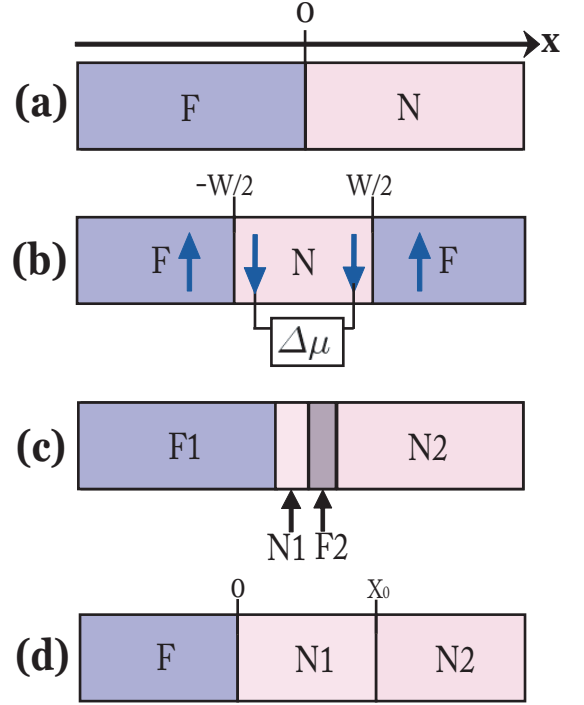}
\caption{Schematic illustration of spin valve system.
(a) a single ferromagnetic layer (F) and nonmagnetic layer (N)
(b) F/N/F structure for parallel spin arrangement between ferromanetic layers. 
A polarized spin current is injected into nonmagnet via ferromanet
perpendicular to the layer.
(c) F1/N1/F2/N2 structure consisting of thin enough  N1 and F2  
and infinitely thick F1 and N2. 
(d) F/N1/N2 structure with a highly doped semiconductor N1 between 
 F and N2.
}
\label{figure1}
\end{center}
\end{figure}

For the case of F/N/F structure as shown in Fig.1(b), 
ECP is an odd function of $x$ \cite{vignale,Alexander}.
In this system, P state or AP state exists since the spin orientations in two F's
are either parallel(P) or antiparallel(AP). 
 In each state, ECP is different and we can obtain the required 
 condition by considering each case.
In AP state, the ECP gradients are derived 
by  
\begin{equation}
\binom{\frac{\partial\mu_{s}}{\partial x}}
{\frac{\partial\mu_{\bar s}}{\partial x}}
=\begin{cases}

\frac{eJ}{\sigma_{\rm F}}\binom{1}{1}+[A_{1}e^{x/l^{\rm F}_{sf}}+B_{1}e^{-x/l^{\rm F}_{sf}}]
\binom{ 1/\sigma_{s} }{ -1/\sigma_{\bar s}  }, \text{$x<-\frac{W}{2}$}\\
\frac{eJ}{\sigma_{\rm N}}[\binom{1}{1}+A_2
{\rm cosh}(\frac{x}{l^{\rm N}_{sf}})\binom{1}{-1}], \text{$-\frac{W}{2}\leq x\leq0$}.\label{ecpsol}
\end{cases}
\end{equation}
where $J$ is the total current density and $\sigma_{\rm F}$ and $\sigma_{\rm N}$ are the total conductivities
in F and N.
In P state, parameters $A_1$, $A_2$, $B_1$ and ${\rm cosh}(\frac{x}{l_{sf}^{\rm N}})$ of Eq.(\ref{ecpsol})
are substituted by 
$A_1'$, $-A_2'$, $B_1'$ and ${\rm sinh}(\frac{x}{l_{sf}^{\rm N}})$.
In the region $x<-\frac{W}{2}$, the condition required to satisfy SSR is
\begin{eqnarray}
\frac{\sigma_{\bar s}}{\sigma_{s}}\bigg\lvert_r=\frac{\sigma_{\bar s}}{\sigma_{s}}\bigg\lvert_{r'}.
\end{eqnarray}
In the region $-\frac{W}{2}\leq x\leq0$, it
becomes simply $\frac{\nabla n_s}{\nabla n_{\bar s}}=-1$.
In two dimensional electron gas,
the conductivity is proportional to the density of states at the constant Fermi level
so that the conductivity $\sigma_{s(\bar s)}$ in N between two ferromagnets 
 satisfies $\sigma_{s(\bar s)}=\sigma_{\rm N}/2$
where $\sigma_{\rm N}$ is the total conductivity in N. 
While $\nabla\mu_{s}\neq\nabla\mu_{\bar s}$ in P state,
$\nabla\mu_{s}=\nabla\mu_{\bar s}$ in AP state
in N. 
That results an unpolarized current flow in AP state and a polarized current flow ($j_s\neq j_{\bar s}$) 
in P state.

Let us take a limiting case of $l^{\rm N}_{sf}\gg l^{\rm F}_{sf}$ such that $\tau_{sf}=\infty$ in N.
In this limit, one can approximate $\mu_{s(\bar s)}\simeq\mu_{0 s(\bar s)}+\Gamma_{s(\bar s)}x$ \cite{Schmidt},
expanding up to the first order in $x$.
The first term $\mu_{0 s(\bar s)}$ denotes ECP with no spin effect.
The constant value of $\frac{\Gamma_{s}}{\Gamma_{\bar s}}$ means 
that SSR is also satisfied in this limit. 
Hence the required condition is trival to satisfy the symmetry relation of XCK
given by Eq.(\ref{syr}). 
The assumption of $\tau_{sf}=\infty$ is valid in several materials such 
as Si-doped GaAs \cite{Kikkawa}. 
Experimental observations in spin valve systems composed of these materials
can be analyzed newly according to SSR.\

The dilute magnetic semiconductor (DMS) has been suggested as an alternative to F due to
large conductivity mismatches between F and N. 
In DMS system, the spin polarization results from the indirect
electron-electron interaction.
The ECPs of DMS/N/DMS structure \cite{Alexander} satisfy Eq.(\ref{ecpsol}), 
except that parameters $A_{i}$, $B_{i}$, $\sigma_{\rm F}$, and $l^{\rm F}_{sf}$ are to be replaced 
respectively by 
$\tilde A_{i}$, $\tilde B_{i}$, $\sigma_{\rm D}$, and $l^{\rm D}_{sf}$
which are determined by the proper boundary conditions.
Hence the investigation of SSR in this system is similar to that of F/N/F structure.

 In F1/N1/F2/N2 multilayer as shown in Fig.1(c), 
electrodes are attached to F1 and N2 to measure the voltage V
across the multilayer. 
In the $i$th layer, ECP is given by \cite{berger}
\begin{eqnarray}
\mu_{is(\bar s)}(x)=C_i+A_{is(\bar s)} e^{\pm x/l_{isf}}
\end{eqnarray}
where the parameters $C_i$ and $A_{is(\bar s)}$ are functions of $\beta$ and 
have different values in different parts of the multilayer.
From the general solution of ECP, 
we see that $\frac{A_{is}}{A_{i\bar s}}=\frac{\beta
+1}{\beta-1}$ in F and $\frac{A_{is}}{A_{i\bar s}}=-1$ in N.
When the ratios $\frac{A_{is}}{A_{i\bar s}}$ at $\boldsymbol r$ and $\boldsymbol r'$
are the same, SSR is always satisfied. \

Let us consider a metallic junction \cite{kimura1}, in which ECPs have the same form
as in F1/N1/F2/N2 structure.
 For the case of Py/Cu/Py system,
 $A_{is(\bar s)}$ and $l_{isf}$ can be obtained in experiments.
For example, when polarized carriers are injected into the Cu wire via the Py pad, 
$A_{is(\bar s)}$ are given by $\mp\mu_i(1\pm\alpha_{\rm Py})/2$ for each part $i$ of the structure
with $l_{{\rm Py} sf}\equiv\lambda_{\rm Py}=2{\rm nm}$ and $\alpha_{\rm Py}=0.2$ \cite{kimura1}. 
Then, $\frac{\nabla n_{s}}{\nabla n_{\bar s}}=-1.5$ at different sites in the present case.\

We consider an F/N1/N2 structure, shown in Fig.1(d), composed of F $(x<0)$, N1 $(0<x<x_0)$
with the conductivity $\sigma_{\rm N1}$, and N2 $(x>x_0)$ with $\sigma_{\rm N2}$.
A highly doped semiconductor N1 is often placed between N2 and F.
The spin densities are given by \cite{Yu}
\begin{equation}
n_{s(\bar s)}=\pm[A_0e^{-x/l_d}+A_1e^{(x-x_0)/l_u}]
\end{equation} 
for $0<x<x_0$
and 
\begin{equation}
n_{s(\bar s)}=\pm[A_2e^{-(x-x_0)/\tilde l_d}]
\end{equation} 
for $x>x_0$
where $l_u$ and $l_d$ are up- and down-stream spin diffusion lengths. 
When $\nabla n^0\lvert_{\boldsymbol r}=0$, 
it is a trivial situation for SSR to be satisfied in undoped and doped systems.\

The SSR for spin currents flowing through the quantum dots 
can also be examined in a similar way.
In magnetic dots, it is interesting to examine the indirect exchange interaction between
conduction electrons and localized spins due to magnetic impurities.
In such magnetic system involving localized d electrons, 
the spin density is written by 
$n_{s(\bar s)}(\boldsymbol r)
=\frac{n_{\rm c}}{2}\mp\frac{9\pi}{2}(\frac{n_c}{\Omega_{\nu}})^2
\frac{J_0}{E_F}\sum_{l}F(2k_F(\vert \boldsymbol r-{\boldsymbol R}_l \vert)<S_{l}^{Z}>$ \cite{Sinha}
where $n_c$, $E_F$, $\boldsymbol R_l$, and $\Omega_{\nu}$
are, respectively, the total number of conduction electrons,
unperturbed Fermi energy,
sites of magnetic ions,
and the volume of a unit cell.
Here, $F(x)=\frac{x{\rm cos}x-{\rm sin}x}{x^4}$.
On this account, the system belongs to an inhomogeneous spin polarized system and
SSR is satisfied for a certainty from the fact that $\frac{\nabla n_s}{\nabla n_{\bar s}}=-1$.\ 

For all these situations satisfying SSR, the properties of the pair correlation function 
can also be known. 
Spin dependent eletron density in a homogeneous spin polarized system
is given, in general, by $n_{s}(\boldsymbol r)=n_s[1-g_{ss}(\boldsymbol r)]+n_{\bar s}[1-g_{s\bar s}(\boldsymbol r)]$ \cite{Yi}.
Hence, Eq.(\ref{key}) can be written by 
\begin{equation}
\frac{n_s\nabla g_{ss}(\boldsymbol r)+n_{\bar s}\nabla g_{s\bar s}(\boldsymbol r)}
 {n_{\bar s}\nabla g_{\bar s \bar s}(\boldsymbol r)+n_{s}\nabla g_{\bar s s}(\boldsymbol r)}\bigg\lvert_{\boldsymbol r}
=\frac{n_s\nabla g_{ss}(\boldsymbol r)+n_{\bar s}\nabla g_{s\bar s}(\boldsymbol r)]}
 {n_{\bar s}\nabla g_{\bar s \bar s}(\boldsymbol r)+n_{s}\nabla g_{\bar s s}(\boldsymbol r)}\bigg\lvert_{\boldsymbol r'}.\label{pair}
\end{equation}
The spin-resolved pair correlation functions including the correlation effect 
can not be evaluated accurately using quantum Monte Carlo algorithm \cite{Palo}.
That is, the accurate condition given by Eq.(\ref{pair})  
can be used as an indicator testing the accuracy of the spin-resolved pair correlation functions.

In summary, we have investigated the symmetry relation of the ``exact'' spin-resolved XCK in broken spin symmetry system.
We have shown the proper cases satisfying SSR in multicomponent structure.
Only by using ECP gradient proportional to the measured spin current,
the properties of XCK can be easily checked.
The proper condition $\frac{\bigtriangledown n_{\bar\sigma}}{\bigtriangledown n_{\sigma}}\big\lvert_{\boldsymbol r}
=\frac{\bigtriangledown n_{\bar\sigma}}{\bigtriangledown n_{\sigma}}\big\lvert_{\boldsymbol r'}$
can be a standard of the correctness of spin-related measurements especially in the trivial situations, 
for example, a homogeneous or nonmagnetic system. 
Hence, we have proposed new method to interpret not only spin current but also the properties of XCK directly
in spintronics. 
We also give the accurate relation of spin-resolved pair distribution functions
which can also be used to test the precision of the pair distribution functions.\

The authors acknowledge the supports in part by the KOSEF(R14-2002-029-01002-0)(IY) and by Korea Research Foundation Grant (KRF-2004-005-C00044)(KSY). 
\renewcommand{\refname}{\large{References}}


\begin{thebibliography}{00}
\bibitem{wolf} S. A. Wolf, et al., Science 294, 1488 (2001).
\bibitem{Kikkawa} J. M. Kikkawa and D. D. Awschalom, Nature (London) 397, 139 (1999).
\bibitem{Yu} Z. G. Yu and M. E. Flatte, Phys. Rev. B {\bf  66}, 235302 (2002).
\bibitem{kimura1} T. Kimura, Y. Otani, and J. Hamrle, 
Phys. Rev. Lett. {\bf  96}, 37201 (2006).
\bibitem{Giuliani} G. Giuliani and G. Vignale, 
{\it Quantum Theory of The Electron Liquid} 
(Cambridge University Press, Cambridge, 2005).
\bibitem{perdew} J. P. Perdew, S. Kurth, A. Zupan, P. Blaha, 
Phys. Rev. Lett. {\bf 82}, 2544 (1999).
\bibitem{phil} P. P. Rushton, D. J. Tozer, and S. J. Clark, 
Phys. Rev. B {\bf 65}, 193106 (2002).
\bibitem{roi} R. Baer and D. Neuhauser, Phys. Rev. Lett. {\bf  94}, 43002 (2005).
\bibitem{julien} J. Toulouse, Phys. Rev. B {\bf  72}, 35117 (2005).
\bibitem{pazini} S. Paziani, S. Moroni, P. Gori-Giorgi, and G. B. Bachelet, Phys. Rev. B {\bf 73}, 155111 (2006).
\bibitem{fran} J. Toulouse, F. Colonna, and A. Savin, 
Phys. Rev. A {\bf  70}, 62505 (2004).
\bibitem{Gori} P. Gori-Giorgi and A. Savin, Phys. Rev. A {\bf  71}, 032513 (2005).
\bibitem{Gonis} A. Gonis, T.C. Schulthess, J. van Ek, and P.E.A. Turchi, 
Phys. Rev. Lett. {\bf  77}, 2981 (1996).
\bibitem{valet} T. Valet and A. Fert, Phys. Rev. B {\bf 48}, 7099 (1993).
\bibitem{vouille} C. Vouille, et al., Phys. Rev. B {\bf 60}, 6710 (1999).
\bibitem{son} P. C. van Son, H. van Kempen, and P. Wyder, 
Phys. Rev. Lett. {\bf 58}, 2271 (1987).
\bibitem{vignale} G. Vignale, Phys. Rev. B {\bf  71}, 125103 (2005).
\bibitem{Alexander} A. Khaetskii, et al, Phys. Rev. B {\bf  71}, 235327 (2005).
\bibitem{Schmidt} G. Schmidt, D. Ferrand, and L. W. Molenkamp, A.T. Filip, B.J. vanWees, Phys. Rev. B {\bf  62}, R4790 (2000).
\bibitem{berger} L. Berger, Phys. Rev. B {\bf 59}, 11465(1999).
\bibitem{Sinha}K. P. Sinha, N. Kumar {\it Interactions in magnetically ordered solids} (Oxford University Press,Oxford, 1980).
\bibitem{Yi} K. S. Yi and J.J Quinn, Phys. Rev. B {\bf  54}, 13398 (1996).  
\bibitem{Palo} S. De Palo, M. Botti, S. Moroni, and Gaetano Senatore, 
Phys. Rev. Lett. {\bf  94}, 226405 (2005).
\end{thebibliography}
\end{document}